\documentclass{ws-ijmpd}

\begin{document}

\markboth{Yongli Ping, Lixin Xu, Hongya Liu and Ying Shao}
{Power-law cosmological solution derived from DGP brane with a brane
tachyon field}
%%%%%%%%%%%%%%%%%%%%% Publisher's Area please ignore %%%%%%%%%%%%%%%
%
\catchline{}{}{}{}{}
%
%%%%%%%%%%%%%%%%%%%%%%%%%%%%%%%%%%%%%%%%%%%%%%%%%%%%%%%%%%%%%%%%%%%%
\title{Power-law cosmological solution derived from DGP brane with a brane tachyon field}

\author{Yongli Ping\footnote{ylping@student.dlut.edu.cn}, Lixin
Xu, Hongya Liu\footnote{hyliu@dlut.edu.cn}}

\address{School of Physics and Optoelectronic Technology,\\ Dalian
University of Technology, Dalian, Liaoning 116024, P.R.China.}
\author{Ying Shao}
\address{Department of physics,\\ Dalian Maritime University,  Dalian, Liaoning 116024,
P.R.China.}

\maketitle

\begin{abstract}
By studying a tachyon field on the DGP brane model, in order to
embed the $4D$ standard Friedmann equation with a brane tachyon
field in $5D$ bulk, the metric of the $5D$ spacetime is presented.
Then, adopting the inverse square potential of tachyon field, we
obtain an expanding universe with power-law on the brane and an
exact $5D$ solution.
\end{abstract}

%\ccode{PACS numbers:04.50.+h; 98.80.-k; 98.80.Cp}

\maketitle

\section{Introduction}
An increasing number of people believe that extra dimensions can be
probed by gravitons and eventually non-standard matter. These models
usually yield the correct Newtonian $(1/r)$-potential at large
distances since the gravitational field is quenched on
sub-millimeter transverse scales. This quenching appears either due
to finite extension of the transverse
dimensions\cite{Arkani-Hamed}\cdash\cite{Kokorelis1} or
sub-millimeter transverse curvature scales derived by negative
cosmological constants.\cite{Randall}\cdash\cite{Abdesselam} A
common feature to these models is that they predict deviations from
the $4D$ Einstein gravity at short distances. Over large distances,
a model which predicts deviations from the standard 4D gravity is
proposed by Dvali, Gabadadze and Porrati
(DGP).\cite{Dvali,Dvali2,Dvali3} There are the brane and bulk
Einstein terms in the action of DGP model. It was shown that the DGP
model allows for an embedding of the standard Friedmann cosmology in
the sense that the cosmological evolution of the background metric
on the brane can entirely be described by the standard Friedmann
equation plus energy conservation on the brane.\cite{Dick,Dick1}
Recently, K. Atazadeh and H. R. Sepangi study the DGP brane with a
scalar field and propose curvature corrections in DGP brane
cosmology.\cite{Atazadeh,Atazadeh1} A comprehensive review on DGP
cosmology is dished up in Ref.~\refcite{Lue}.

The observable universe is presently undergoing an accelerating
expansion basing on the observations of Type Ia
supernovae.\cite{Riess,Perlmutter,Spergel} It is possible that such
an accelerated expansion could be the result of a modification to
the Einstein-Hilbert action.\cite{Deffayet}\cdash\cite{Lue1}
Recently, it has been proposed that accelerated expansion of
universe is driven by a tachyon
field.\cite{Choudhury}\cdash\cite{Calcagni} This field is derived
from string-brane physics and describes the lowest energy level of
an unstable Dp-brane or that of a brane-antibrane system.\cite{Sen}
And the notable characteristic of tachyon field is that the tachyon
field has a negative squared mass.\cite{Gibbons}\cdash\cite{A. de la
Macorra} Moreover, there are several papers about the brane-world
model with a tachyon
field.\cite{Papantonopoulos}\cdash\cite{Mukohyama}

In Ref.~\refcite{Atazadeh}, DGP brane cosmology with a brane scalar
field is introduced. Now, we use a tachyon field take the place of
the scalar field on the brane. In this paper, it is found that
standard Friedmann equation with a brane tachyon field embeds in
$5D$ bulk; then, by adopting a common potential of tachyon field, we
obtain the metric which satisfies a power-law expansion on the
brane. At the same time, an exact $5D$ solution is derived.

\section{DGP model with a brane tachyon field}
The action for the DGP model with a scalar field on the brane is
written as\cite{Atazadeh}
\begin{equation}
S=\frac{M^3_5}{2}\int{d^5x\sqrt{-g}\mathcal{R}}
+\int{d^4x\sqrt{-q}\left[\frac{M^2_{pl}}{2}R-\frac{1}{2}(\nabla{\phi})^2-V(\phi)\right]}
+S_m[q_{\mu\nu},\psi_m],\label{action1}
\end{equation}
where the first term in (\ref{action1}) describes the
Einstein-Hilbert action in $5D$ bulk for a five-dimensional metric
$g_{AB}$ with Ricci scalar $\mathcal {R}$. And the second term is
the Einstein-Hilbert action for the induced metric $q_{\mu\nu}$ on
the brane with a scalar field $\phi$ and $R$ is the scalar field of
brane. $M_5$ is the Planck mass in five dimensions and $M_{pl}$ is
the induced $4D$ Planck mass. The last term $S_m$ is the matter
action on the brane with matter field $\psi$. The metric
$q_{\mu\nu}$ is induced from the bulk metric $g_{AB}$ via
\begin{equation}
q_{\mu\nu}=\delta_{\mu}^A\delta_{\nu}^Bg_{AB}.
\end{equation}

Now assuming a tachyon field instead of the scalar field on the
brane, the action is rewritten as

\begin{equation}
S=\frac{M^3_5}{2}\int{d^5x\sqrt{-g}\mathcal{R}}
+\int{d^4x\sqrt{-q}\left[\frac{M^2_{pl}}{2}R-V(T)\sqrt{1+\partial_{\mu}{T}\partial^{\mu}{T}}\right]}
+S_m[q_{\mu\nu},\psi_m].\label{action}
\end{equation}
From the action (\ref{action}), the Einstein equations are derived
as
\begin{equation}
M^3_5\left(\mathcal {R}_{AB}-\frac{1}{2}g_{AB}\mathcal{R}\right)
+M^2_{pl}\delta^{\mu}_A\delta^{\nu}_B\left(R_{\mu\nu}-\frac{1}{2}q_{\mu\nu}R\right)\delta(y)
=\delta^{\mu}_A\delta^{\nu}_B\left(T_{\mu\nu}+\mathcal
{T}_{\mu\nu}\right)\delta(y),\label{e-e}
\end{equation}
here, $T_{\mu\nu}$ and $\mathcal {T}_{\mu\nu}$ are the
energy-momentum tensor in the tachyon field and matter field
respectively. The corresponding junction conditions\cite{Dick}
become

\begin{eqnarray}
\lim_{\epsilon\rightarrow+0}[K_{\mu\nu}]^{+}_{-}
&=&\frac{1}{M^3_5}\left(T_{\mu\nu}+\mathcal
{T}_{\mu\nu}-\frac{1}{3}q_{\mu\nu}q^{\alpha\beta}(T_{\alpha\beta}+\mathcal
{T}_{\alpha\beta})\right)\Big |_{y=0} \nonumber\\
&-&\frac{M^2_{pl}}{M^3_5}\left(R_{\mu\nu}-\frac{1}{6}q_{\mu\nu}q^{\alpha\beta}R_{\alpha\beta}\right)\Big
|_{y=0}.\label{junction}
\end{eqnarray}
From the Lagrangian of the tachyon field, the density and pressure
are given
\begin{eqnarray}
\rho_T&=&\frac{V(T)}{\sqrt{1-\dot{T}^2}}\label{rho-T},\\
p_{T}&=&-V(T)\sqrt{1-\dot{T}^2}.\label{pT}
\end{eqnarray}
Upon variation of the action, the equation of the motion for the
tachyon field can be written
\begin{equation}
\frac{V(T)\ddot{T}}{1-\dot{T}^2}+3\frac{\dot{a}}{a}V(T)\dot{T}+\frac{dV(T)}{dT}=0
\label{field-V}.
\end{equation}
The form of the line element is written as in the brane gravity
\begin{equation}
ds^2=-n^2(y,t)dt^2+a^2(y,t)\gamma_{ij}dx^{i}dx^{j}+b^2(y,t)dy^2,
\end{equation}
here, $\gamma_{ij}$ is a maximally symmetric $3D$ metric where
$k=0,\pm1$ parameterizes the spatial curvature. Now, we will follow
the work in  Ref.~\refcite{Dick}, adopting the Gaussian normal
system gauge $b^2(y,t)=1$, the Einstein tensors in the bulk are
\begin{eqnarray}
G_{00}&=&3n^2\left(\frac{\dot{a}^2}{n^2a^2}-\frac{a'^2}{a^2}+\frac{k}{a^2}\right)-3n^2\frac{a''}{a}\label{G00},\\
G_{ij}&=&\left(\frac{a'^2}{a^2}-\frac{\dot{a}^2}{n^2a^2}-\frac{k}{a^2}\right)g_{ij}\nonumber\\
&+&2\left(\frac{a''}{a}
+\frac{n'a'}{na}-\frac{\ddot{a}}{n^2a}+\frac{\dot{n}\dot{a}}{n^3a}\right)g_{ij}+\frac{n''}{n}g_{ij}\label{Gij},\\
G_{05}&=&3\left(\frac{n'\dot{a}}{na}-\frac{\dot{a}'}{a}\right)\label{G05},\\
G_{55}&=&3\left(\frac{a'^2}{a^2}-\frac{\dot{a}^2}{n^2a^2}-\frac{k}{a^2}\right)
+3\left(\frac{n'a'}{na}+\frac{\dot{n}\dot{a}}{n^3a}-\frac{\ddot{a}}{n^2a}\right)\label{G55}.
\end{eqnarray}

The matching condition (\ref{junction}) for the perfect fluid on the
brane
\begin{equation}
\mathcal {T}_{00}=\rho,\ \ \ \mathcal {T}_{11}=\mathcal
{T}_{22}=\mathcal {T}_{33}=p
\end{equation}
read
\begin{eqnarray}
\lim_{\epsilon\rightarrow+0}[n']^{+}_{-}&=&\frac{n}{3M^3_5}\left(2(\rho+\rho_T)+3(p+p_{T})\right)\Big|_{y=0}\nonumber\\
&+&\frac{M^2_{pl}}{M^3_5}2n\left(\frac{\ddot{a}}{n^2a}-\frac{\dot{a}^2}{2n^2a^2}
-\frac{\dot{n}\dot{a}}{n^3a}-\frac{k}{2a^2}\right)\Big|_{y=0},\\
\lim_{\epsilon\rightarrow+0}[a']^{+}_{-}&=&\frac{M^2_{pl}}{M^3_5}\left(\frac{\dot{a}^2}{n^2a}+\frac{k}{a}\right)\Big|_{y=0}
-\frac{(\rho+\rho_{T})a}{3M^3_5}\Big|_{y=0}\label{m-condition}.
\end{eqnarray}
Since $T_{05}+\mathcal {T}_{05}=0$ from (\ref{G05}), we get
$n'/n=\dot{a}'/\dot{a}$. Then, in order to simplify, adopting the
gauge $n(0,t)=1$, it is obtained
\begin{equation}
n(y,t)=\frac{\dot{a}(y,t)}{\dot{a}(0,t)}\label{n}.
\end{equation}
Therefore, from (\ref{G00}), (\ref{G55}) and (\ref{m-condition}), we
have
\begin{eqnarray}
\lim_{\epsilon\rightarrow0}[a']^+_-(t)&=&\frac{M^2_{pl}}{M^3_5}\left[\frac{\dot{a}^2(0,t)}{a(0,t)}
+\frac{k}{a(0,t)}\right]\bigg|_{y=0}-\frac{(\rho+\rho_T)a(0,t)}{3M^3_5}\Big|_{y=0},\\
I^+&=&\left[\dot{a}^2(0,t)-a'^2(y,t)+k\right]a^2(y,t)\Big|_{y>0}, \\
I^-&=&\left[\dot{a}^2(0,t)-a'^2(y,t)+k\right]a^2(y,t)\Big|_{y<0}.
\end{eqnarray}
By taking $I^+=I^-$, the embedding of standard Friedmann cosmology
with a tachyon field is given as follows:
\begin{eqnarray}
&&\frac{\dot{a}^2(0,t)+k}{a^2(0,t)}=\frac{\rho+\rho_T}{3M^2_{pl}}\label{s-F},\\
&&I=\left[\dot{a}^2(0,t)-a'^2(y,t)+k\right]a^2(y,t)\label{I}.
\end{eqnarray}
From (\ref{n}) and (\ref{I}), we can obtain the components of metric
for $I=0$, that is
\begin{eqnarray}
a(y,t)&=&a(0,t)+\sqrt{\dot{a}^2(0,t)+k}y,\label{a(y,t)}\\
n(y,t)&=&1+\frac{\ddot{a}(0,t)}{\sqrt{\dot{a}^2(0,t)+k}}y.\label{n(y,t)}
\end{eqnarray}
Therefore, given a specific $a(0,t)$, the exact solution of $5D$
metric is determined by equations (\ref{a(y,t)},\ref{n(y,t)}). Then,
it is presented the mathematic configuration of $5D$ brane world.
This metric is similar to the one in space-time-matter
theory\cite{Liu,Liu1} and brane model\cite{Liu2}. There are some
link between them.\cite{Ping}

\section{The power-law cosmological solution for a given $V(\phi)$}
From (\ref{field-V}) and (\ref{s-F}) for the spatially flat FRW
metric, we have
\begin{equation}
3H^2_0=\frac{1}{M^2_{pl}}(\rho+\rho_T),\label{F-H}\\
\end{equation}
and
\begin{equation}
\frac{\ddot{T}}{1-\dot{T}^2}+3H_0\dot{T}+\frac{1}{V(T)}\frac{dV(T)}{dT}=0\label{field-H}
\end{equation}
where $H_0={\dot{a}(0,t)}/{a(0,t)}$. Then, we assume the tachyon
field dominates the universe, i.e. $\rho=0$. Considering the tachyon
field, the potential of this field is arbitrary. But most adopt the
inverse square potential. We utilize this potential as
\begin{equation}
V_{T}=2nM^2_{pl}(1-\frac{2}{3n})^{1/2}T^{-2}
\end{equation}
with $ n\geq2/3$, which is shown in Ref.~\refcite{Padmanabhan} in
order to obtain an accelerated expansion of the universe driven by
tachyonic matter. Substituting this potential into (\ref{F-H}) and
(\ref{field-H}) with $M^2_{pl}=1$, in the brane the evolution of the
tachyon field is
\begin{equation}
T\propto\left(\frac{2}{3n}\right)^{1/2}t,\label{T}
\end{equation}
and the evolution of the scale factor is obtained
\begin{equation}
a(0,t){\propto}t^{n}.\label{a1}
\end{equation}
Therefore, the deceleration parameter $q$ is rewritten as
\begin{equation}
q=-1+\frac{1}{n}.
\end{equation}
The deceleration parameter $q$ is plotted with $n$.
\begin{figure}
\begin{center}
\includegraphics[angle=0,width=3.0in]{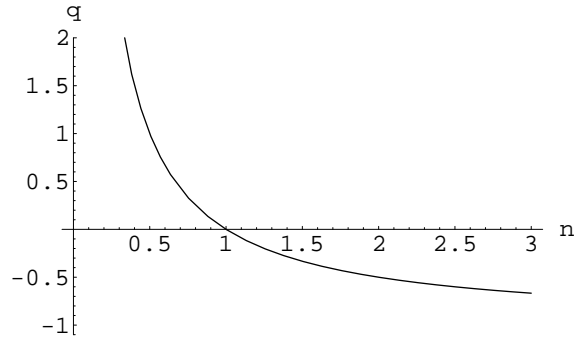}
\end{center}
\caption{Evolution of the deceleration parameter $q$ vs.
$n$}\label{q}
\end{figure}
From the Fig.{\ref{q}}, we find that when $2/3\leq{n}<1$, this
predicts deceleration on the brane; when $n=1$, there is an uniform
speed expansion on the brane; and when $n>1$, an accelerated
universe is obtained on the brane. Substituting (\ref{a1}) into
(\ref{a(y,t)}) and (\ref{n(y,t)}),for $k=0$ we derive
\begin{equation}
a(y,t)=C\left[t^{n}+nt^{n-1}y\right]\label{a2}
\end{equation}
and
\begin{equation}
n(y,t)=C\left[1+(n-1)\frac{y}{t}\right],\label{n2}
\end{equation}
where $C$ is a constant. When $y=0$, Eq. (\ref{a2}) and (\ref{n2})
return to (\ref{a1}) and $n(0,t)=1$. There appears coordinate
singularities on the space-like hypercone $y=\pm{t/(n-1)}$. This is
presumably a consequence of the fact that the orthogonal geodesics
emerging from the brane (which we used to set up $b^2=1$) do not
cover the full five-dimensional spacetime. Therefore, the exact
solution is
\begin{equation}
ds^2=-[1+(n-1)\frac{y}{t}]dt^2+[t^n+nt^{n-1}y]\gamma_{ij}dx^idx^j+dy^2.
\end{equation}
This solution can lead to arbitrarily velocity expansion with $n$ on
the brane. For getting the uniform speed expansion universe $n=1$,
this solution is rewritten as
\begin{equation}
ds^2=-dt^2+[t+y]\gamma_{ij}dx^idx^j+dy^2,
\end{equation}
which is a critical situation.

To simplify, we choose a more simple form of potential as
\begin{equation}
V_{T}=M^2T^{-2}.
\end{equation}
Substituting this potential into (\ref{F-H}) and (\ref{field-H}),
the evolution of the tachyon field on the brane is
\begin{equation}
T\propto\left(\frac{4}{2+\sqrt{9M^2+4}}\right)^{1/2}t\label{T1}.
\end{equation}
Therefore, when $M^2=1$, we obtain the tachyon field is
\begin{equation}
T\propto\left(\frac{4}{2+\sqrt{13}}\right)^{1/2}t,\label{T}
\end{equation}
and the evolution of the scale factor is
\begin{equation}
a(0,t){\propto}t^{(1/3+\sqrt{13}/6)}.\label{a11}
\end{equation}
Substituting (\ref{a1}) into (\ref{a(y,t)}) and (\ref{n(y,t)}),for
$k=0$ we derive
\begin{equation}
a(y,t)=C\left[t^{({1}/{3}+{\sqrt{13}}/{6})}+({\frac{1}{3}+\frac{\sqrt{13}}{6})yt^{(\sqrt{13}/6-2/3)}}\right]\label{a21}
\end{equation}
and
\begin{equation}
n(y,t)=C\left[1+(\frac{\sqrt{13}}{6}-\frac{2}{3})\frac{y}{t}\right],\label{n21}
\end{equation}
where $C$ is a constant. However, when
$a(0,t){\propto}t^{(1/3+\sqrt{13}/6)}$,
${(1/3+\sqrt{13}/6)}\approx0.93<1$. So, the deceleration parameter
$q>0$. Therefore, we get an exact $5D$ cosmological solution which
leadS to decelerated expansion on the brane.

\section{Conclusion}
In this paper, we consider the DGP model with a brane tachyon field.
In order to make the standard Friedmann cosmology with the tachyon
field embed in the $5D$ bulk, the metric of $5D$ spacetime is
derived. In this form of the metric, we find if the scale factor $a$
is given the exact solution of the $5D$ bulk will be obtained. Then,
it is known that if the potential of the tachyon field is given, the
scale factor $a$ will be described. For a general form of the
inverse square potential, an arbitrarily velocity expanding universe
with power-law is obtained on the brane. Meanwhile the correspond
$5D$ solution is derived. The velocity of expansion is related to
$n$. If $2/3\leq{n}<1$, the universe is decelerating expansion; if
$n=1$, the universe is uniform speed expansion; and if $n>1$, the
accelerated expansion universe is obtained. For example, giving
specific potential $V_{T}=M^2T^{-2}$, when $M^2=1$, we have the
decelerated expansion universe and the exact solution of $5D$ bulk.

\section*{Acknowledgments}
This work was supported by NSF (10573003), NSF (10647110), NSF
(10703001), NBRP (2003CB716300) of P. R. China and DUT 893321.

\end{document}